\documentclass[aps,twocolumn,prb,superscriptaddress,showpacs]{revtex4}
\usepackage{amsmath,amssymb,latexsym}
\usepackage[dvips]{graphicx}

\begin{document}

\title{Raman spectra of BN-nanotubes:
Ab-initio and bond-polarizability model calculations}

\author{Ludger Wirtz}
\altaffiliation[New address: ]{
IEMN (CNRS-UMR 8520), B. P. 60069, 59652 Villeneuve d'Ascq Cedex, France}
\affiliation{Department of Material Physics, University of the Basque
Country, Centro Mixto CSIC-UPV, and Donostia International Physics 
Center (DIPC),
Po.~Manuel de Lardizabal 4, 20018 Donostia-San Sebasti\'an, Spain}

\author{Michele Lazzeri}
\affiliation{Institut de Min\'eralogie et de Physique des Milieux Condens\'es, 
4 Place Jussieu, 75252 Paris cedex 05, France}

\author{Francesco Mauri}
\affiliation{Institut de Min\'eralogie et de Physique des Milieux Condens\'es, 
4 Place Jussieu, 75252 Paris cedex 05, France}

\author{Angel Rubio}
\affiliation{Department of Material Physics, University of the Basque
Country, Centro Mixto CSIC-UPV, and Donostia International Physics 
Center (DIPC),
Po.~Manuel de Lardizabal 4, 20018 Donostia-San Sebasti\'an, Spain}

\date{\today}

\begin{abstract}
We present {\it ab-initio} calculations of the non-resonant Raman spectra 
of zigzag and armchair BN nanotubes. In comparison, we implement
a generalized bond-polarizability model where the parameters are
extracted from first-principles calculations of the polarizability
tensor of a BN sheet.
For light-polarization along the tube-axis, the agreement between
model and {\it ab-initio} spectra is almost perfect. For perpendicular
polarization, depolarization effects have to be included 
in the model in order to reproduce the {\it ab-initio} Raman intensities.
\end{abstract}

\maketitle

Besides its success in the characterization of a large range of materials
\cite{cardona},
Raman spectroscopy has also developed into an invaluable tool for the 
characterization of nanotubes. Since the first characterization of 
(disordered) carbon nanotube (CNT) samples \cite{rao97}, the technique has
been refined, including, e.g., polarized Raman studies of aligned nanotubes 
\cite{rao00} and isolated tubes \cite{due00}.
On the theoretical side, non-resonant Raman intensities of CNTs have been 
calculated within the bond-polarizability model \cite{sai98,dressbook1}.
The empirical parameters of this model are adapted to fit experimental
Raman intensities of fullerenes and hydrocarbons.
However, the transferability of the parameters and the quantitative
performance in nanotubes, in particular distinguishing between
metallic and semiconducting tubes, is still not clear.

In this communication, we report on the Raman spectra of boron nitride
nanotubes (BNNTs) \cite{rub94,cho95}. Recently, synthesis of BNNTs 
in gram quantities has been reported \cite{lee01}. Their characterization 
through Raman and infrared spectroscopy is expected to play an important 
role. However, due to difficulties with the sample
purification no experimental data on contamination-free samples has
been reported. Ab-initio ~\cite{wir03} and 
empirical~\cite{san02,pop03}
phonon calculations have determined the
position of the peaks in the spectra. However, due to 
missing bond-polarizability parameters for BN, the Raman
intensities have been so far addressed 
using the model bond-polarizability parameters of carbon \cite{pop03}.
Only the intensities of high-frequency modes was presented
as it was argued that the intensity of
low frequency modes are very sensitive to the
bond-polarizability values \cite{pop03}.
Here, we derive the polarizability parameters for BN sp$^2$ bonds from
a single hexagonal BN sheet by calculating 
the polarizability-tensor and its variation under deformation.
We compare the resulting spectra for BNNTs with full {\it ab-initio}
calculations. We derive conclusions about
the general applicability of the bond polarizability model
for semiconducting CNTs.

In non resonant first-order Raman spectra, peaks appear at 
the frequencies $\omega_\nu$ of the $\nu$ optical phonon
with null wave vector with intensities $I^\nu$ which in
the Placzek approximation \cite{cardona} are
\begin{equation}
I^\nu \propto \left|{\bf e}_i \cdot {\bf A}^\nu \cdot {\bf e}_s \right|^2
\frac{1}{\omega_\nu}(n_\nu + 1) \; . 
\end{equation}
Here ${\bf e}_i$ (${\bf e}_s$) is the polarization of the incident (scattered)
light and $n_\nu = [\exp(\hbar \omega_\nu/k_B T)-1]^{-1}$ with $T$ being
the temperature.
The Raman tensor ${\bf A}^\nu$ is
\begin{equation}
A_{ij}^\nu = \sum_\gamma B_{ij}^{k\gamma} 
\frac{ w_{k\gamma}^\nu}{\sqrt{M_\gamma}},
\end{equation}
where $w_{k\gamma}^\nu$ is the $k$th Cartesian component of atom
$\gamma$ of the $\nu$th orthonormal vibrational eigenvector and $M_\gamma$
is the atomic mass.
\begin{equation}
B_{ij}^{k\gamma} = \frac{\partial^3 \mathcal{E}}{\partial E_i \partial E_j
\partial u_{k\gamma}} = \frac{\partial \alpha_{ij}}{\partial u_{k\gamma}},
\label{devtens}
\end{equation}
where $\mathcal{E}$ is the total energy of the unit cell, ${\bf E}$ is 
a uniform electric field and $u_{k\gamma}$ are atomic displacements.
This is equivalent to the change of the electronic polarizability
of a unit cell, $\alpha_{ij} = \Omega \chi_{ij}$ (where $\Omega$ is the
unit cell volume and $\chi_{ij}$ the electric susceptibility), upon 
the displacement $u_{k\gamma}$.
The phonon frequencies and eigenvectors \cite{wir03} are determined by 
density functional perturbation theory~\cite{bar01}
as implemented in Ref. \cite{abinit}. 
For the determination of the derivative tensor $B_{ij}^{k\gamma}$ we proceed
in two ways: i) we calculate it from first principles using the
approach of Ref. \onlinecite{laz03},
ii.) we develop a generalized bond-polarizability model.

The basic assumption of the bond polarizability model 
\cite{wol41,cardona,uma01} is that the total polarizability can be  modeled
in terms of single bond contributions. Each bond is assigned a longitudinal
polarizability, $\alpha_l$, and a polarization perpendicular to the bond,
$\alpha_p$. Thus, the polarizability contribution $\alpha_{ij}^b$ of a 
particular bond $b$ is
\begin{equation}
\alpha_{ij}^b = \frac{1}{3}(2\alpha_p + \alpha_l)\delta_{ij}
+ (\alpha_l - \alpha_p)\left(\hat{R}_i\hat{R}_j - \frac{1}{3}\delta_{ij}\right),
\label{special}
\end{equation}
where $\hat{R}$ is a unit vector along the bond.
The second assumption is that the bond polarizabilities
only depend on the bond length $R$. This allows the
calculation of the derivative with respect to atomic displacement,
$\partial \alpha_{ij}^b/\partial u_{k\gamma}$,
in terms of four parameters $\alpha_l(R)$, $\alpha_p(R)$, $\alpha_l'(R)$,
and $\alpha_p'(R)$ (see, e.g. Ref. \onlinecite{uma01}).
The use of only one perpendicular parameter $\alpha_p$ implicitly assumes 
cylindrical symmetry of the bonds. That can be justified in a sp$^3$ bonding
environment. However, in the highly anisotropic environment 
in a sheet of sp$^2$ bonded carbon or BN
and the corresponding nanotubes this assumption seems hardly justified.
In our model we therefore define a generalized 
polarizability with an in-plane ($\alpha_{pi}$) and out-of-plane value 
($\alpha_{po}$) of $\alpha_{p}$.

With the larger set of parameters, the polarizability tensor takes on 
the more general form
\begin{equation}
\alpha_{ij}^b = \alpha_l \hat{R}_i\hat{R}_j
            + \alpha_{pi} \hat{S}_i\hat{S}_j
            + \alpha_{po} \hat{T}_i\hat{T}_j,
\label{general}
\end{equation}
where $\hat{\bf S}$ is a unit-vector pointing
perpendicular to the bond in plane, and $\hat{\bf T}$ pointing
perpendicular to the bond out of plane.
(In the case of $\alpha_{pi} = \alpha_{po}$, Eq.~(\ref{general}) simplifies to 
Eq. (\ref{special}) due to the relation 
$ \hat{S}_i\hat{S}_j + \hat{T}_i\hat{T}_j = \delta_{ij} - \hat{R}_i\hat{R}_j$.)
For the derivative tensor (of a single bond), we obtain
\begin{eqnarray}
\frac{\partial \alpha_{ij}^b}{\partial u_{k\gamma}}  & = &
   \alpha_l' \hat{R}_i\hat{R}_j\hat{R}_k
  +\alpha_l ((\partial_k\hat{R}_i)\hat{R}_j + \hat{R}_i(\partial_k\hat{R}_j))
\nonumber \\
& + & \alpha_{pi}' \hat{S}_i\hat{S}_j\hat{R}_k
  +\alpha_{pi} ((\partial_k\hat{S}_i)\hat{S}_j + \hat{S}_i(\partial_k\hat{S}_j))
\nonumber \\
& + & \alpha_{po}' \hat{T}_i\hat{T}_j\hat{R}_k
  +\alpha_{po} ((\partial_k\hat{T}_i)\hat{T}_j + \hat{T}_i(\partial_k\hat{T}_j)).
\label{der2}
\end{eqnarray}
The total derivative tensor $B_{ij}^{k\gamma}$ is then just the
sum over all $\partial \alpha_{ij}^b/\partial u_{k\gamma}$ of all
bonds of the system.
The orientation of the plane at the 
position of a particular atom is thereby defined by the three nearest neighbor 
atoms.

\begin{table}
\begin{tabular}{|l|c|c|c|c|c|} \hline
         &  R ({\AA})  &  $\alpha_l$ ({\AA}$^3$) &  $\alpha_p$ ({\AA}$^3$) & $\alpha_l'$ ({\AA}$^2$) & $\alpha_p'$ ({\AA}$^2$) \\
\hline
         &             &                         &  $\alpha_{pi}$: 0.28    &                         & $\alpha_{pi}'$: 6.60    \\
\raisebox{1.5ex}[-1.5ex]{BN-sheet} &  \raisebox{1.5ex}[-1.5ex]{1.44} & \raisebox{1.5ex}[-1.5ex]{3.31} & 
$\alpha_{po}$: 0.44 & \raisebox{1.5ex}[-1.5ex]{1.03} & $\alpha_{po}'$: 0.77 \\
\hline
c-BN     &  1.56 & 1.58 & 0.42 & 4.22 & 0.90 \\
\hline
diamond  &  1.53 & 1.69 & 0.71 & 7.43 & 0.37 \\
\hline
\end{tabular}
\caption{Parameters of the bond polarizability model extracted from
{\it ab-initio} calculations (see text).}
\label{alphatable}
\end{table}

\begin{figure}[btp]
 \centering
   \includegraphics[draft=false,keepaspectratio=true,clip,%
                    width=1\linewidth]%
                    {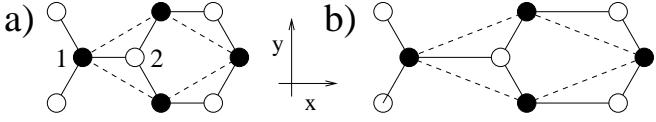}
\caption[FIG1]{Unit cell (marked by dashed line) of BN-sheet for the 
calculation of
the bond polarizability parameters: a) equilibrium geometry,
b) geometry with one bond elongated.}
\label{cellfig} 
\end{figure}

In order to determine the six parameters of our
 model, we perform {\it ab-initio} calculations of the polarizability
tensor $\alpha_{ij}$ of a unit-cell of a single BN-sheet
\cite{distnote,claus1dim} (see Fig.~\ref{cellfig} a).
The geometry of the system leads to the relations 
$\alpha_{xx} = \alpha_{yy} = (3/2)(\alpha_l + \alpha_{pi})$ and
$\alpha_{zz} = 3\alpha_{po}$ (with the z-axis perpendicular to the sheet).
Displacing atom 2 in $y$-direction yields the relation
$\partial \alpha_{xx}/\partial u_{2y} = (3/4)(\alpha_l' + \alpha_{pi}')
+ (3/2)(\alpha_l + \alpha_{pi})/R $.
Finally, by changing the geometry of the unit-cell such that one
bond is elongated while the other two bond-lengths and all the bond angles 
are kept constant (see Fig.~\ref{cellfig} b), we
extract the derivatives of the bond polarizabilities: 
$\alpha_{l}'=\alpha_{xx}'$, $\alpha_{pi}'=\alpha_{yy}'$, 
and $\alpha_{po}'=\alpha_{zz}'$.
The resulting parameters are displayed in Tab. I
and compared to the parameters we calculated for cubic BN and diamond.
The longitudinal bond polarizability
$\alpha_l$ is considerably larger than $\alpha_p$ which can be intuitively
explained as a consequence of the ``enhanced mobility" of the electrons along 
the bond. For the sheet, the perpendicular polarizabilities clearly
display different values in the in-plane and out-of-plane directions.
Without the added flexibility of different parameters, the bond-polarizability
 model
would lead to inconsistencies in the description of $\alpha_{ij}$ and
its derivatives. In the sheet, $\alpha_l$ is about twice as large as in
cubic BN (c-BN) due to the additional contribution of the $\pi$
electrons to the longitudinal polarizability. Comparison of c-BN with
the isoelectronic diamond shows a slightly higher polarizability of
the C-C bond.

\begin{figure}[btp]
 \centering
   \includegraphics[draft=false,keepaspectratio=true,clip,%
                    width=1\linewidth]%
                    {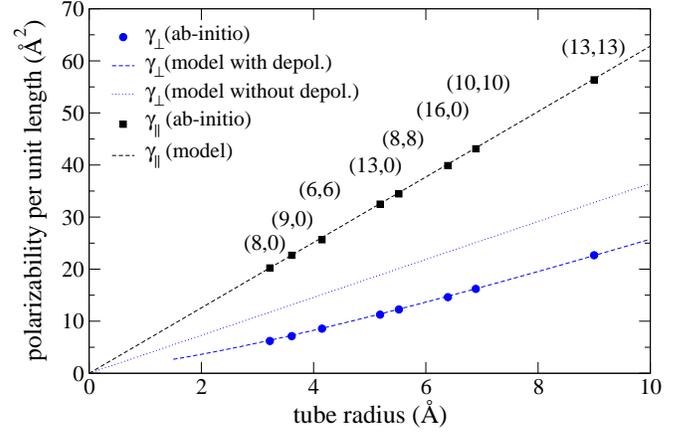}
\caption[FIG2]{Perpendicular ($\gamma_\perp$) and longitudinal
($\gamma_\parallel$) polarizabilities per unit length
of different BN-nanotubes:
{\it ab-initio} and our generalized
bond polarizability model.
The influence of depolarization can be seen for
$\gamma_\perp$.}
\label{tubepolfig} 
\end{figure}

As a first application of the generalized bond-polarizability
 model, we present in 
Fig.~\ref{tubepolfig} the polarizability $\gamma$ (per unit length) of 
different BNNTs \cite{claus2dim}.
For the polarizability along the tube axis (z-direction), the model
(Eq.~(\ref{general}))
agrees almost perfectly with our {\it ab-initio} calculations.
The polarizability is proportional to the number of bonds in the
unit-cell which is proportional to the tube radius.
For the perpendicular direction,
the model calculations overestimate the {\it ab-initio} values
considerably.
This discrepancy demonstrates the importance of {\em depolarization effects}
in the perpendicular direction: 
due to the inhomogeneity of the charge distribution in this direction,
an external field induces local fields which counteract the external field
and thereby reduce the overall polarizability. The size of this
effect can be estimated from a simple model. Imagine a dielectric hollow 
cylinder of radius $R$ (measured at the mid-point between inner
and outer wall) and thickness $d$. 
The dielectric constant in tangential 
direction, $\epsilon_\parallel = (d + 4\pi\beta_\parallel)/d$, 
is different from the dielectric constant
in radial direction, $\epsilon_\perp = d/(d-4\pi\beta_\perp)$.
Here, $\beta_\parallel$ and $\beta_\perp$ are the polarizabilities
per unit area of the BN-sheet which are extracted from the bulk
calculation~\cite{claus1dim}.
The polarizability $\gamma$ per unit length of the cylinder due to an external 
homogeneous electric field perpendicular to the tube axis is \cite{henr96}
\begin{equation}
\gamma(R) = - \frac{1}{2} \left(R+\frac{d}{2}\right)^2 
\frac{(\epsilon_\parallel \epsilon_\perp -1)
(1 - \Theta^{2\nu})}{(\sqrt{\epsilon_\parallel \epsilon_\perp} - 1)^2
\Theta^{2\nu} - (\sqrt{\epsilon_\parallel \epsilon_\perp} + 1)^2},
\label{term1}
\end{equation}
with $\Theta = (R-d/2)/(R+d/2)$ and 
$\nu = \sqrt{\epsilon_\parallel/\epsilon_\perp}$.
In the limit $R/d \rightarrow \infty$, the polarizability in Eq.~(\ref{term1}) 
displays a linear dependence on the radius:
$\gamma(R) \rightarrow \gamma_0(R) - \delta$,
where $\gamma_0(R) = \pi (\beta_\parallel + \beta_\perp) R$.
This corresponds to the polarizability without depolarization effects
and coincides with the undamped model curve for $\gamma_{\perp}$ (dotted line
in Fig.~\ref{tubepolfig}).
 
The depolarization effects are introduced into our
model by multiplying the undamped model curve for the perpendicular 
polarizability 
with the ``damping'' factor $\Gamma(R) = \gamma(R)/\gamma_0(R)$. This 
factor depends on the cylinder thickness $d$. The value $d=3$\AA,
which corresponds approximately to the full width of the charge-density
of a BN-sheet, leads to an almost
perfect agreement between model and {\it ab-initio} calculations.\cite{footd}

To compute Raman intensities we make the further assumption:
\begin{equation}
B_{ij}^{k\gamma} = \frac{\partial (\Gamma_{ij} \alpha_{ij})}
{\partial u_{k\gamma}}
\simeq \Gamma_{ij} \frac{ \partial \alpha_{ij}}{\partial u_{k\gamma}}, 
\label{assump}
\end{equation}
where
$\partial \alpha_{ij}/\partial u_{k\gamma}$ is constructed according
to Eq.~\ref{der2}. We assume here that to first order the atomic displacement does 
not change the depolarization. For $i=j=3$, i.e., for incoming
and scattered light polarized along the tube axis, $\Gamma_{ij} = 1$,
otherwise $\Gamma_{ij} = \Gamma(R)$.

\begin{figure}[btp]
 \centering
   \includegraphics[draft=false,keepaspectratio=true,clip,%
                    width=1\linewidth]%
                    {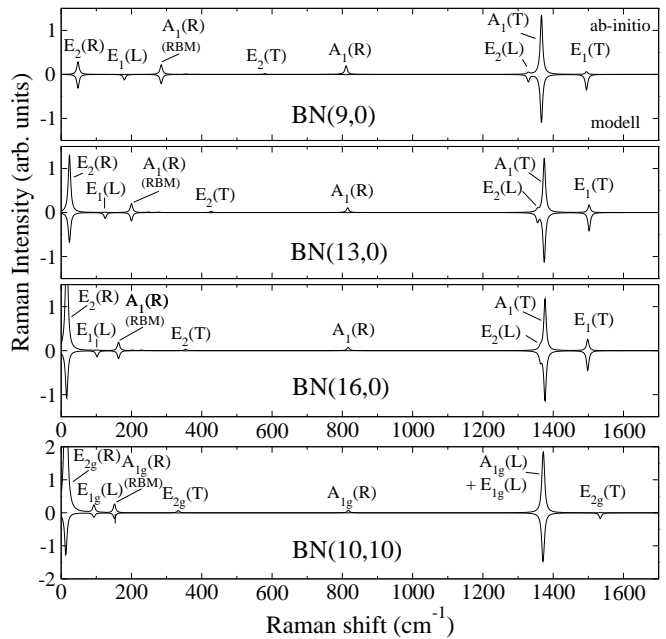}
\caption[FIG3]{Raman spectrum for different BN tubes:
Comparison of {\it ab-initio} calculations (positive axis) 
with the bond-polarization model (inverted axis).
Symmetry assignment follows Ref.~\onlinecite{dam01}.
The letters R,T,L denote
the character of the corresponding phonon oscillation: radial, transverse,
or longitudinal (see Ref.~\onlinecite{wir03}).} 
\label{ram1fig} 
\end{figure}

In Fig.~\ref{ram1fig} we present the {\it ab-initio} and model
Raman spectra for the (9,0), (13,0), and (16,0) zigzag BN nanotubes and 
a (10,10) armchair tube. The latter two have diameters (12.8 \AA
and 13.8 \AA) in the range of 
experimentally produced BN tubes~\cite{cho95,lee01}. The spectra
are averaged over the polarization of the incoming light and
scattered light. We first discuss the
spectra of the zigzag tubes.
The peaks below 700 cm$^{-1}$ are due to low frequency phonon modes which
are derived from the acoustic modes of the sheet and whose frequencies scale
inversely proportional to the tube diameter (except for the E$_2$(R) mode 
which scales with the inverse square of the diameter) \cite{wir03}. 
The E$_2(R)$ mode gets quite intense with increasing tube diameter,
 but its frequency is so low that it will be hard
to distinguish it from the strong Raleigh-scattering peak in experiments.
The E$_1$(L) peak has almost vanishing intensity in the {\it ab-initio}
spectrum and is overestimated in the model. 
The radial breathing mode (RBM) yields a clear peak which should be easily 
detectable in Raman measurements of BNNTs just as in the case of CNTs.
Both {\it ab-initio} and model calculations yield a similar intensity for
this peak. 
The high-frequency modes above 700 cm$^{-1}$ are derived from the optical 
modes of the sheet and change weakly with diameter. 
The A$_1$(R) mode at 810 cm$^{-1}$ gives
a small contribution which might be detectable. The intensity
decreases, however, with increasing diameter. The model only yields
a vanishingly small intensity for this peak. At 1370 cm$^{-1}$ a clear
signal is given by the A$_1$(T) mode which has very similar intensity
both in model and {\it ab-initio} calculations. The small side peak
at slightly lower frequency is due to the E$_2$(L) mode.
The E$_1$(T) peak at 1480 cm$^{-1}$ is gaining intensity with increasing
tube radius. The overall Raman spectrum for a (10,10) armchair tube exhibit
similar trends.

\begin{figure}[btp]
 \centering
   \includegraphics[draft=false,keepaspectratio=true,clip,%
                    width=1\linewidth]%
                    {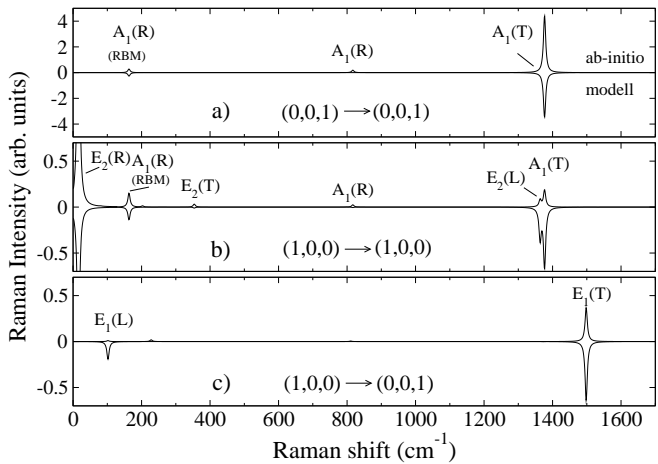}
\caption[FIG4]{Raman spectrum of a BN(16,0) tube for different
light polarizations ${\bf e}_i \rightarrow  {\bf e}_f$.
Tube oriented along (001).}
\label{rampolfig} 
\end{figure}

In Fig.~\ref{rampolfig} we show for the (16,0) tube the dependence of the 
intensities on the light polarizations.
If both ${\bf e}_i$ and ${\bf e}_f$ point along
the tube axis (Fig.~\ref{rampolfig} a), only the A$_1$ modes
are visible and described well by the model (except the 810 cm$^{-1}$ mode). 
This coincides with the finding that for the polarizability along
the tube axis, depolarization does not play a role \cite{mar03}.
The E modes are only visible if at least one of ${\bf e}_i$ and ${\bf e}_f$  
has a component perpendicular to the tube axis (Fig.~\ref{rampolfig} b and c).
The bond polarizability model reproduces these peaks, but tends to overestimate
the E modes. The inclusion of depolarization effects is 
absolutely mandatory. Without depolarization, the model overestimates the
Raman intensities for perpendicular polarization by about a factor of 15.
Other discrepancies are due to the assumption
in Eq.~(\ref{assump}).

In conclusion, we implemented the bond polarizability model
for BN nanotubes with parameters taken from {\it ab-initio} calculations
and under inclusion of depolarization effects.
Going beyond previous models for graphitic systems,
our calculations yield different parameters 
for the in-plane and out-of-plane perpendicular polarizabilities.
Good agreement between model and {\it ab-initio} calculations
of the non-resonant Raman spectra of BN nanotubes 
is obtained for light polarization along the tube axis. For perpendicular
polarization, the inclusion of depolarization effects leads to a 
reasonable agreement between model and {\it ab-initio} spectra.
The model is implemented for single-wall BN tubes but can be extended
to multi-wall tubes if the strength of the depolarization effects is
modeled accordingly. A similar bond-polarizability model 
can also be developed for the non-resonant Raman spectra of semiconducting 
carbon NTs. However, due to the metallic behavior, a bond-polarizability
model is not applicable to the graphene sheet. Consequently, the modeling
of the polarizability of semiconducting tubes is very sensitive to the 
band-structure \cite{benedict}, in particular to the band-gap which
depends on the radius and chirality of the tubes.
This work was supported by EU
Network of Excellence NANOQUANTA (NMP4-CT-2004-500198)
and Spanish-MCyT.
Calculations were performed at IDRIS and CEPBA supercomputer centers.



\begin{thebibliography}{99}
\bibitem{distnote}
We use a periodic supercell with an intersheet distance of 12 a.u.
We use Troullier-Martins pseudopotentials.
The two dimensional Brillouin zone
is sampled by a 12$\times$12 Monckhorst-Pack grid.
Local density approximation is used.
\bibitem{claus1dim}
The polarizability perpendicular to the sheet
is calculated through the 1-dimensional Clausius-Mossotti relation
$\alpha_{\perp} = \alpha_{zz} 
= \frac{\Omega}{4\pi}\frac{\epsilon_{zz}-1}{\epsilon_{zz}}$
while the parallel polarizability obeys the relation
$\alpha_{xx} =  \frac{\Omega}{4\pi}(\epsilon_{xx}-1)$.
$\Omega$ is the unit-cell volume.
We checked numerically that the thus obtained
$\alpha_{zz}$ and $\alpha_{xx}$ are independent of the inter-sheet
distance. The polarizability per unit area, $\beta$, is
$\beta = \alpha/A$, where $A$ is the unit-cell area in the
x-y-plane.
\bibitem{claus2dim}
The perpendicular tube polarizability 
is extracted from the bulk dielectric tensor through the 2-dimensional
Clausius-Mossotti relation
$\alpha_{\perp} = 
\frac{2\Omega}{4\pi}\frac{\epsilon_{\perp}-1}{\epsilon_{\perp}+1}$.
We checked numerically that the calculated 
$\alpha_{\perp}$ is
approximately independent of the inter-tube distance.
\bibitem{footd} The dependence on the empirical parameter $d$ is weak:
For the (16,0) tube, changing $d$ from 2.2 \AA to 4 \AA gives rise
to a change of 2 \% in the perpendicular polarizability.
\end{thebibliography}
\end{document}